\documentclass[a4paper,fleqn,usenatbib]{mn2e}

\usepackage{ae,aecompl}

\usepackage{graphicx}	
\usepackage{amsmath}	
\usepackage{amssymb}	

\usepackage{mathtools,cuted}
\usepackage{perpage}

\usepackage{color}
\usepackage{aas_macros}
\usepackage{hyperref}

\title[Cosmological magnetic braking and SMBH]{Cosmological magnetic braking and the formation of high-redshift, super-massive black holes}

\author[Kanhaiya L. Pandey et al.]{
Kanhaiya L. Pandey,$^{1}$\thanks{E-mail: kanhaiya.pandey@iiap.res.in (KLP); sethi@rri.res.in (SKS); ratra@phys.ksu.edu (BR)}
Shiv K. Sethi,$^{2 \star}$
and Bharat Ratra$^{3 \star}$
\\
$^{1}$Indian Institute of Astrophysics, II Block, Koramangala, Bangalore 560 034, India\\
$^{2}$Raman Research Institute, Sadashivanagar, Bangalore 560080, India\\
$^{3}$Department of Physics, Kansas State University, 116 Cardwell Hall, Manhattan, Kansas 66506, USA
}

\date{Accepted XXX. Received YYY; in original form ZZZ}

\pubyear{2015}

\begin{document}
\label{firstpage}
\pagerange{\pageref{firstpage}--\pageref{lastpage}}
\maketitle

\begin{abstract}

We study the effect of magnetic braking due to a  primordial magnetic field in the context of the formation of massive ($\gtrsim 10^{4} M_\odot$) direct collapse black holes (DCBHs) at high redshifts. Under the assumption of axial symmetry, we  analytically compute  the effect of magnetic braking on the angular momentum of gas collapsing into the potential well of  massive dark matter haloes  ($\simeq 10^{7-9} M_\odot$)  which are spun up by  gravitational tidal torques. We find that a primordial magnetic field of strength $B_0\simeq 0.1$~nG (comoving)  can remove   the initial angular momentum gained by the in-falling gas due to   tidal torques, thus  significantly lowering the angular momentum barrier to the formation of DCBHs. These magnetic field strengths are consistent with the bounds on primordial fields from astrophysical and cosmological measurements and they are large enough to seed observed galactic magnetic fields. 

\end{abstract}




\section{Introduction}

Recent discoveries of high luminosity ($L > 10^{47}$~erg s$^{-1}$) quasars at high redshifts ($z\simeq 6-7$) suggest that some  $10^{8-9} M_\odot$ super-massive black holes (SMBHs) were present when the Universe was less than a Gyr old \citep{2011Natur.474..616M,2015Natur.518..512W,2018arXiv180706055S,2018arXiv181012310W}. One leading explanation for the formation of such massive black holes at high redshifts is the direct collapse black hole (DCBH) formation model. This model suggests that rapid collapse of metal-free primordial gas into the gravitational potential of a sufficiently massive dark matter halo ($M_h \gtrsim 10^{8}M_\odot$), with virial temperature $T_{\rm vir}\gtrsim 10^4$ K, can form a direct collapse black hole of mass $10^{4-6}M_\odot$ which can further accrete gas from the surrounding medium and turn into a SMBH \citep{2002ApJ...569..558O,2003ApJ...596...34B,2005ApJ...633..624V,2006MNRAS.370..289B}. For the collapsing gas to become a black hole the angular momentum barrier has to be smaller than the Schwarzschild radius corresponding to the mass of the collapsing gas. \citet{1988ApJ...329..589R} has studied how collapsing halo density perturbations get torqued up through tidal interactions with the surrounding spatial inhomogeneity density field as they evolve and subsequently acquire a net angular momentum. For the collapsing gas to form a black hole there has to be an efficient mechanism to transfer this angular momentum from the collapsing cloud to the surrounding medium on the dynamical timescale. 

In this paper we follow \citet{1995Princeton}\footnote{Available at https://www.phys.ksu.edu/personal/ratra/.} and consider the possibility of using magnetic braking caused by a primordial cosmological magnetic field to reduce the angular momentum barrier for such collapsing halos. The Universe is known to be magnetized on all scales probed so-far, from small scales such as planets to large scales such as galaxies and cluster of galaxies. Recent observational evidence suggests that the intergalactic medium and voids could also harbor magnetic fields of strength $> 10^{-16 \pm 1}$ gauss (comoving) coherent over Mpc scales \citep{2010Sci...328...73N}.\footnote{It is also possible, although thought less probable, that these observations instead indicate stronger magnetic fields coherent over smaller length scales. There also has been a debate about whether plasma instabilities might be able to explain these observations. Recent discussions of these matters are given by \citet{2014ApJ...787...49S}, \citet{2015ApJ...814...20F}, \citet{2016A&A...585A.132K}, \citet{2016PhRvD..94h3005A}, \citet{2017ApJ...835..288A}, \citet{2018ApJ...857...43V}, \citet{2018ApJ...859...45S}, \citet{2018ApJ...868...87B}, and \citet{2019ApJ...870...17Y}, from which earlier developments may be traced.}
 A plausible explanation for the existence of such large-scale coherent magnetic fields is that they are of primordial cosmological origin. Several mechanisms have been proposed to produce large-scale primordial magnetic fields of strength up to a few nanoGauss (comoving) during the inflationary era and later during early universe phase transitions \citep[and references therein]{2012SSRv..166...37W,2013A&ARv..21...62D,2016RPPh...79g6901S,2018arXiv181011876K}.\footnote{Inflation is currently the most promising scenario for generating a cosmological magnetic field \citep{1992ApJ...391L...1R}. To generate a large enough magnetic field while allowing inflation to proceed, the abelian gauge field, coupled to the inflaton dilaton scalar field, must be strongly coupled during inflation \citep[][available at https://www.phys.ksu.edu/personal/ratra/]{1991Caltech}. It is not yet known if this semiclassical magnetogenesis mechanism can be accommodated in a consistent quantum mechanical setting. For recent discussions see \citet{2016PhRvD..94j3510C}, \citet{2017PhRvD..95h3509V}, \citet{2017JCAP...06..035M}, \citet{2018CQGra..35l4003C}, \citet{2017PhRvD..96h3511S}, \citet{2018PhRvD..97h3503S}, \citet{2018PhRvD..98f3534S}, \citet{2018JCAP...10..040S}, \citet{2018JCAP...10..023B}, \citet{2018arXiv181003478C}, and references therein. Other cosmological magnetogenesis mechanisms are discussed by \citet{2017PhLB..768...46A}, \citet{2017JCAP...04..030K}, \cite{2018PhLB..785..399F}, \citet{2018PhRvL.121c1102C}, \citet{2018arXiv180602505S}, \citet{2018JHEP...11..039B}, and references therein.}

If such primordial magnetic fields existed they can be significantly amplified \citep[][and references therein]{1970AuJPh..23..731P} by flux-freezing of these fields inside  gas collapsing  under the influence of  dark matter  potential wells and so can play an important role in early structure formation \citep{2005PhRvD..72j3003G,2010PhRvD..82h3005K,2010ApJ...721..615S,2012ApJ...748...27P,2013ApJ...762...15P,2015MNRAS.451.1692P,2016MNRAS.456L..69M}.\footnote{A cosmological magnetic field can be observationally constrained through the effects it has on structure formation. It can also be bounded by how it affects big bang nucleosynthesis and the cosmic microwave background radiation. Each of these three probes limit the magnetic field strength to $< {\rm few} \times 10^{-9}$ G (comoving) on a Mpc scale. For discussions of these bounds, see \citet{2007PhRvD..75b3002K}, \citet{2010PhRvD..82h3005K}, \citet{2016A&A...594A..19P}, \citet{2015PhRvD..92l3509A}, \citet{2017PhRvD..95f3506Z}, \citet{2018CQGra..35l4004P}, \citet{2018PhRvD..97j3525Y}, and \citet{2018arXiv181200730M}, through which earlier developments may be traced.}

In this paper we focus on the idea originally proposed in the context of star formation by \citet{1960EvHT...184.....E} and \citet{1965QJRAS...6..265M} and later elaborated by \citet{1979ApJ...230..204M} (referred to as MP79 henceforth):  a frozen-in magnetic field can brake the rotation of a collapsing cloud by trying to force it to corotate with the surrounding medium, with the resulting Alfv{\'e}n waves transporting away some of the angular momenta from the cloud. \citet{1995Princeton} used time scale estimates to show that magnetic braking by a cosmological magnetic field of strength needed to explain galactic magnetic fields in the anisotropic collapse and differential rotation amplification scenario \citep{1970AuJPh..23..731P,1988plap.work..197K} would efficaciously remove angular momentum during the cosmological formation of primordial Pop III stars and black holes (in the DCBH formation scenario). In our more complete dynamical analysis here we solve for the final angular momentum acquired by the halo by simultaneously taking into account both the tidal torque due to the surrounding density field which spins up the halo as well as the effect of magnetic braking owing to  the presence of a cosmological magnetic field. 

Our analysis shows that the presence of a cosmological magnetic field of sufficient current strength $\sim 0.1$ nG coherent over an Mpc scale could in fact play an important role in removing angular momentum and so allow the formation of high-redshift massive seed black holes in the DCBH scenario. 

We give a brief description of the DCBH scenario for SMBH formation and the associated angular momentum barrier issue in the following section (Sec.~2). In Secs.~3 and 4 we provide a detailed description of our analysis and results. The last section (Sec.~5) summarizes the main results of this paper.

\section{Angular momentum transfer during  DCBH formation}

To make the DCBH model work one needs to devise a mechanism that allows the gas to rapidly collapse without fragmenting while efficiently shedding angular momentum. For these conditions to be met the collapsing gas must maintain a high temperature ($\gtrsim 10^4$)~K so that the sound speed is large enough to allow rapid in-fall of gas. To ensure near isothermal collapse at ($\gtrsim 10^4$)~K, the  collapsing gas has to be almost metal free to avoid fragmentation \citep{2006MNRAS.370..289B,2010MNRAS.402.1249S,2013ApJ...774..149C,2015MNRAS.450.4411C}. This kind of rapid gas collapse is expected in relatively massive dark matter halos that have  virial temperatures $\gtrsim 10^4$~K. Unlike the case for stellar mass black holes,  the  collapsing gas remains optically thin to the radiation  produced during the formation of high mass ($\sim 10^{4-5} M_\odot$) black holes, as the density remains low even close to the Schwarzschild radius. Under these conditions, gas fragmentation into subclumps is almost completely inhibited \citep{2002ApJ...571...30S,2005ApJ...626..627O,2014MNRAS.442.2036D}. As a result, the collapsing gas continues to lose thermal pressure as it collapses and can directly  form a massive black hole without going through a stellar phase. However, as the density of the collapsing gas increases it starts forming H$_2$ molecules which can rapidly cool the gas and lower the temperature to about $\sim 200$ K, which could cause  fragmentation  as the  Jeans mass is smaller  at  lower temperatures. This can be avoided by invoking mechanisms to destroy H$_2$ molecules, such as photo-dissociation of H$_2$ molecules due to UV flux from nearby galaxies, heating due to magnetic field decay, or heating due to accreting primordial blackholes \citep{2003ApJ...596...34B,2008ApJ...686..801O,2008MNRAS.391.1961D,2010MNRAS.402.1249S,2010ApJ...721..615S,2018JApA...39....9P}.

For the collapsing gas to form a black hole there must also be an efficient mechanism to transfer the angular momentum of the collapsing cloud to the surrounding medium on the dynamical timescale. For this to happen the collapsing gas must be able to lose angular momentum at early stages of the collapse. For a typical  order of magnitude estimate, during the collapse this would require the angular speed $v_\phi(t)$ obey \citep{2004PhRvD..69j4016D}
\begin{equation}
v_\phi(t) \lesssim \frac{c R_S}{\ r(t)},
\label{eq:vphi_limit1}
\end{equation}
where $c$ is the speed of light in vacuum, $R_S$ is the Schwarzschild radius corresponding to the mass of the collapsing cloud, and $r(t)$ is the radius of the cloud. The above equation can be expressed as the scaling relation at the virialization time
\begin{equation}
v_\phi(t_{\rm vir}) \lesssim \left( \frac{M_t}{10^8 M_\odot} \right)^{2/3} 3 \times 10^{-4} \ {\rm km}\ {\rm s}^{-1},
\label{eq:vphi_limit2}
\end{equation}
where $M_t$ is the total mass (dark matter halo + baryons/gas) of the object.  Here we assume that the underlying baryonic component has the same angular speed as the dark matter halo at the initial time. If, by the time of virialization, some  mechanism can reduce the angular speed  of the collapsing cloud below the limit given by Eq.~(\ref{eq:vphi_limit2}), the cloud must collapse into a black hole provided  that it does not gain any angular momentum from the surroundings during the course of the collapse and provided that the cooling time scale remains smaller than the dynamical timescale. 

\section{Methodology}

\subsection{The initial conditions}

To model the spin-up of a collapsing halo due to tidal torquing and  its spin down owing to cosmological magnetic field braking, we  choose  suitable initial conditions described in the previous section.  We  use the spherical tophat collapse model to study the collapse  of gas  into  potential wells of three dark matter halo with (CDM) masses $M_h = 10^7$, $10^8$, and $10^9 M_\odot$, and virial temperature $T_{\rm vir} \gtrsim 10^4$~K. More specifically, the initial conditions we use are: initial density contrast  $\bar\delta_i \sim 0.04$ (averaged over the overdensity under consideration) at $z_i = 1000$ which corresponds to a nearly $6 \sigma$ fluctuation for the $\Lambda$CDM model with virialization redshift $z_{\rm vir} \sim  16$ and $z_{\rm max} \sim  25$ ($z_{\rm max}$ is the redshift at which the dark matter halo radius is the maximum). 

We use the  spherical top-hat model in the redshift range  $500 \gtrsim z \gtrsim 5$. In this model, the comoving radius of the dark matter halo is  given by the parametric equations
        \begin{eqnarray}
        \label{eq:top-hat}
            r(\theta) &=& \frac{r_{\rm max}}{2} (1-\cos\theta) \\
            t(\theta) &=& B (\theta - \sin\theta) ,
        \end{eqnarray}
where $r_{\rm max}=(3/5) r_i \bar\delta_i^{-1}$ is the maximum radius achieved by the dark matter halo and $B = (3 t_i/4) (3/5 \bar\delta_i)^{3/2}$. The dark matter halo initial radius $r_i$ at initial time $t_i$ is
        \begin{equation}
            r_i = \left(\frac{3M_h}{4 \pi \rho_{\rm ext,i} (1+\bar\delta_i)}\right)^{1/3} ,
        \end{equation}
where the background nonrelativistic matter density $\rho_{\rm ext}$ of the Universe at redshift $z_i$ is
        \begin{equation}
            \rho_{\rm ext,i}= \frac{3 H_0^2 \Omega_m(1+z_i)^3}{8 \pi G}  .
        \end{equation}
Here $H_0$ is the Hubble constant and the current value of the nonrelativistic matter density parameter $\Omega_m = \Omega_b + \Omega_c$, where $\Omega_b$ and $\Omega_c$ are the current values of the baryonic matter and CDM density parameters. For the background cosmology we assume the standard spatially-flat $\Lambda$CDM model \citep{1984ApJ...284..439P} which is consistent with most observations.\footnote{We note however that current data cannot rule out mildly closed spatial hypersurfaces \citep{2018ApJ...864...80O,2018arXiv180100213P} or mild dark energy dynamics \citep{2018arXiv180205571O,2018arXiv180903598P}.} 

We assume that the gas cloud of baryon mass $M_b = M_h\Omega_b/\Omega_c$ is dynamically co-evolving with the dark matter halo  and hence has the same radius as that of the dark matter halo until virialization. 

\subsection{Tidal torque due to the surrounding density field}

The collapsing halo gains angular momentum through tidal interactions with the surrounding density field \citep[see][and references therein]{1988ApJ...329..589R}. The torque $N_{\rm tid}$ due to these tidal interactions can be expressed as \citep{1988ApJ...329..589R} 
        \begin{equation}
            N_{\rm tid}(\theta) = \left(\frac{2}{9}\right)^{2/3} \frac{\tau_0 }{\bar{\delta_i}} \frac{(1-\cos\theta)^2}{(\theta-\sin\theta)^{4/3}} \frac{f_2(\theta)}{f_1(\theta) - f_2(\theta)},
           \end{equation}
where
        \begin{eqnarray}
            f_1(\theta) & = & 16 - 16 \cos\theta + \sin^2\theta - 9 \theta \sin\theta , \\
            f_2(\theta) & = & 12 - 12 \cos\theta + 3 \sin^2\theta - 9 \theta \sin\theta ,
           \end{eqnarray}
and $\tau_0$ is the value of tidal torque at the initial time $t_i$.  The value of $\tau_0$ is taken from  \citet{1988ApJ...329..589R}, scaled for different values  of halo mass $M_h$. We assume that the tidal torque switches off once the halo reaches its maximum radius. 

\begin{figure*}
\begin{center}
\includegraphics[width=0.40\textwidth]{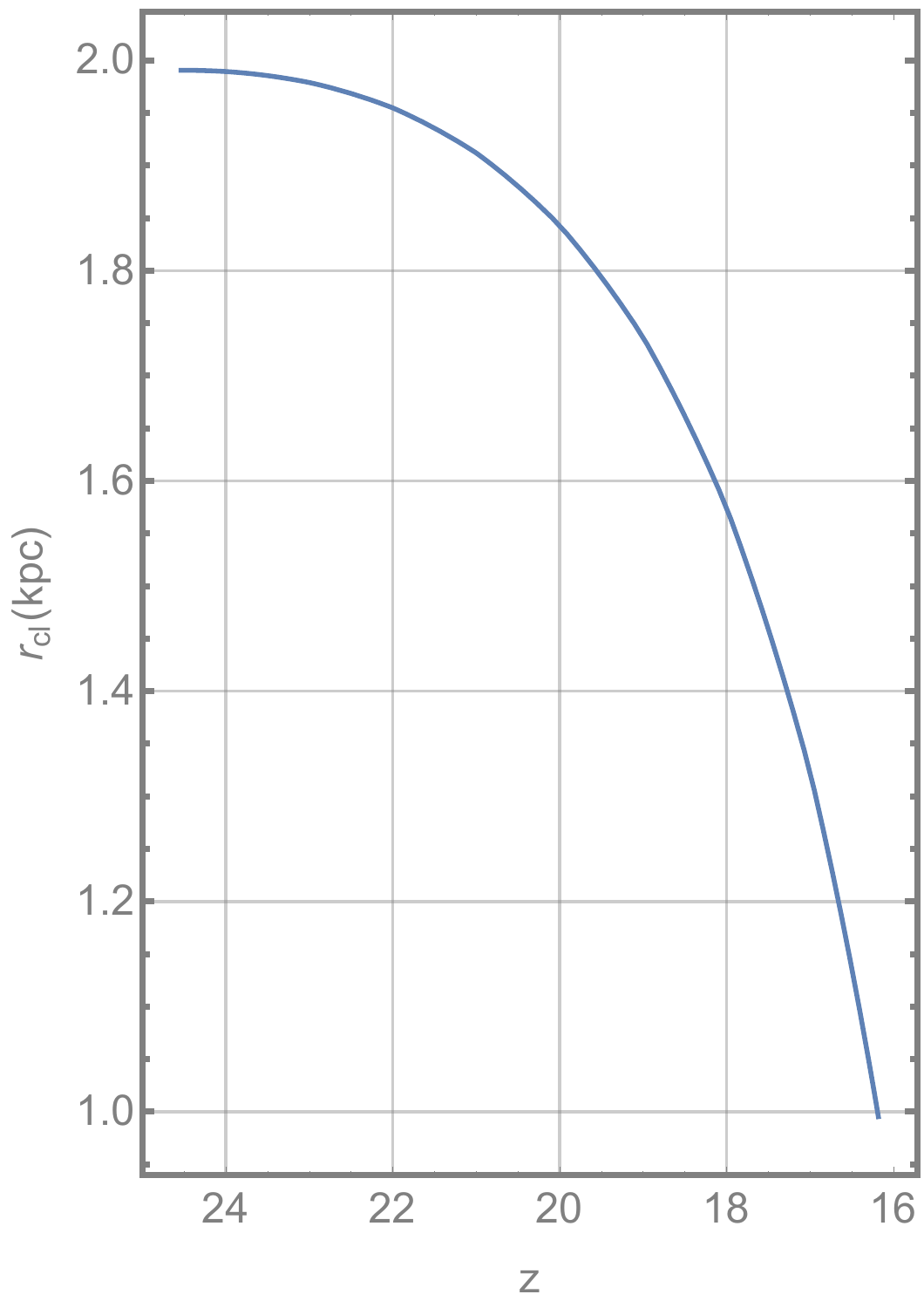}
\hspace{1cm}
\includegraphics[width=0.41\textwidth]{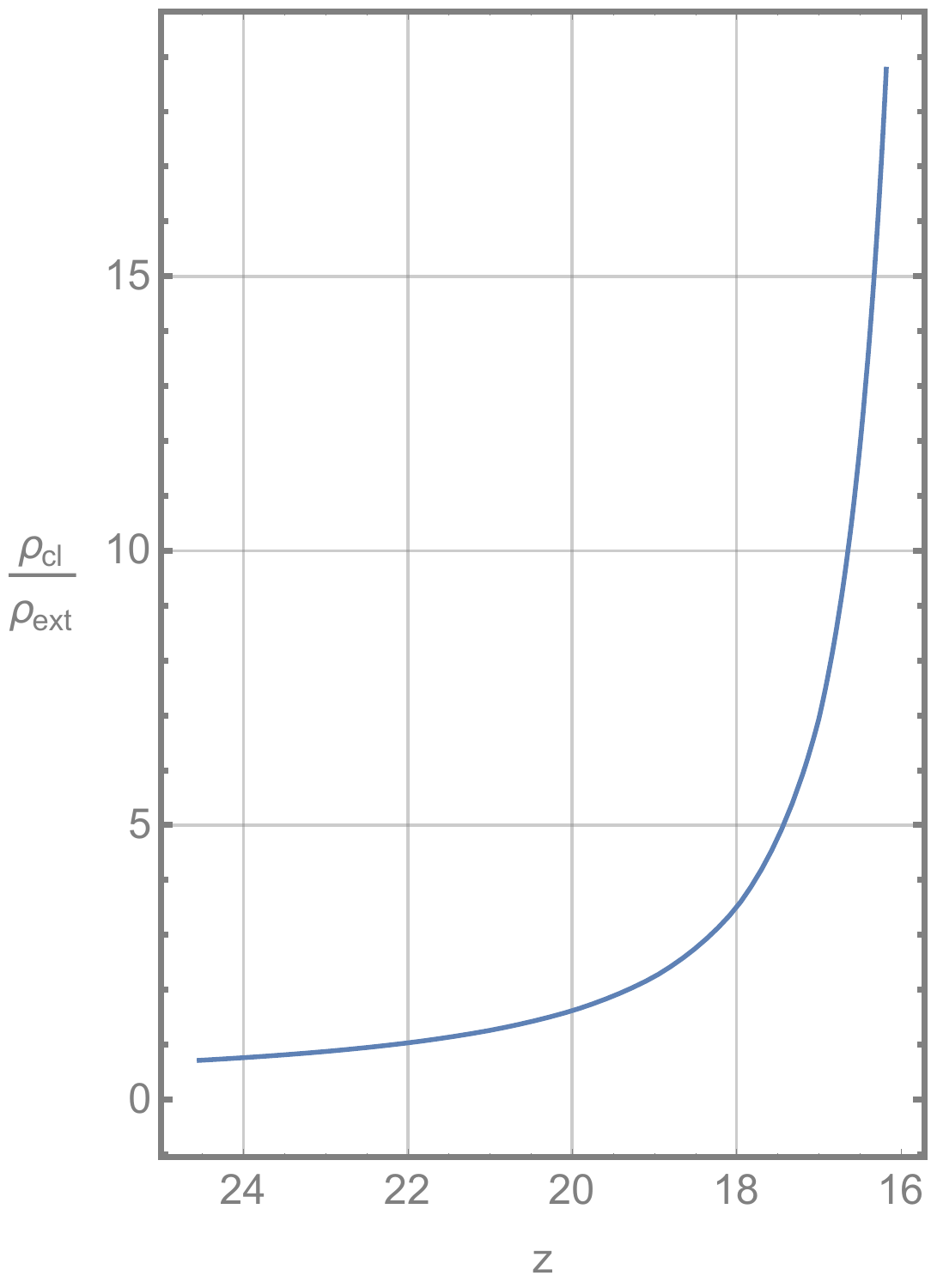}
\includegraphics[width=0.40\textwidth]{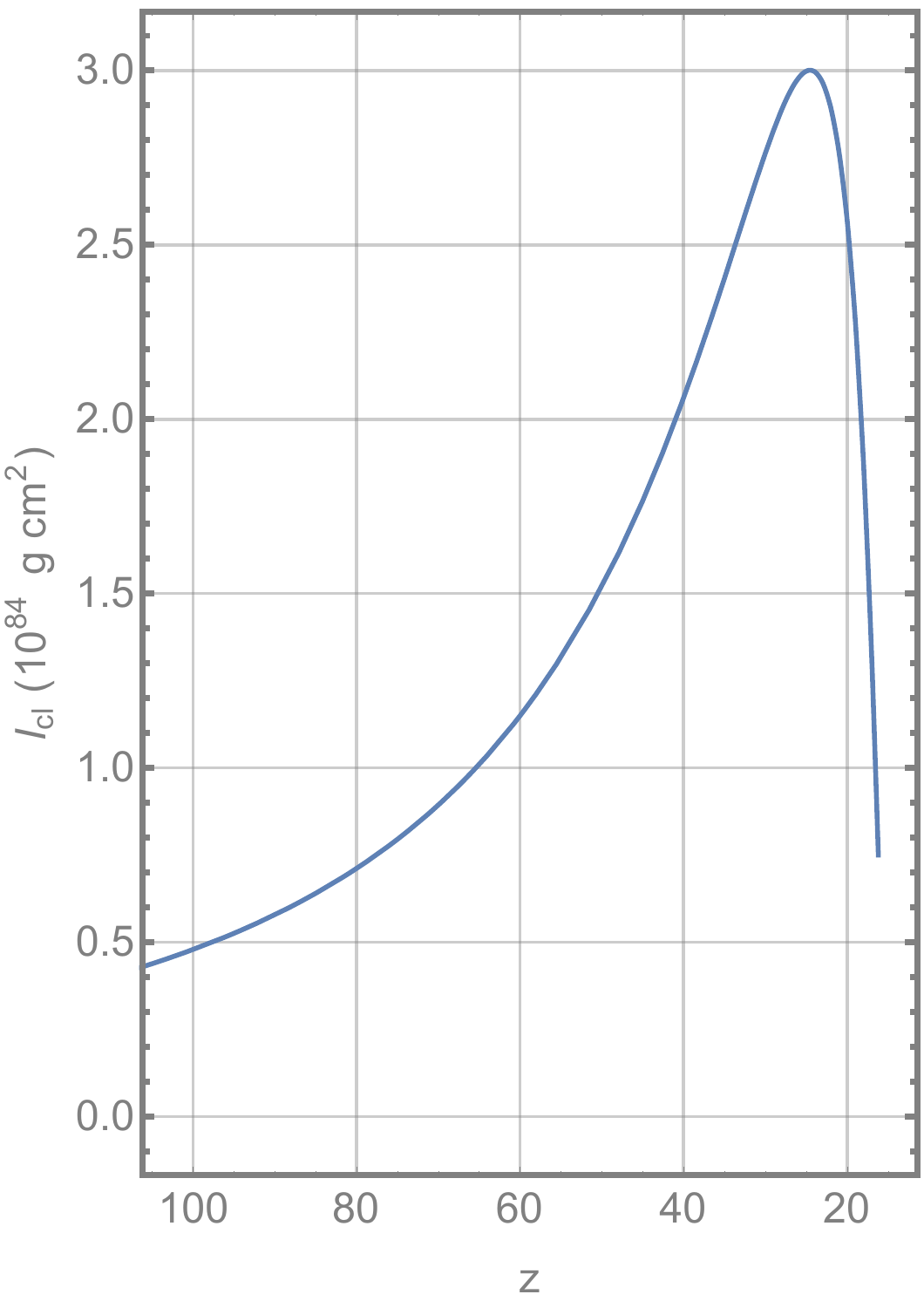}
\hspace{1cm}
\includegraphics[width=0.40\textwidth]{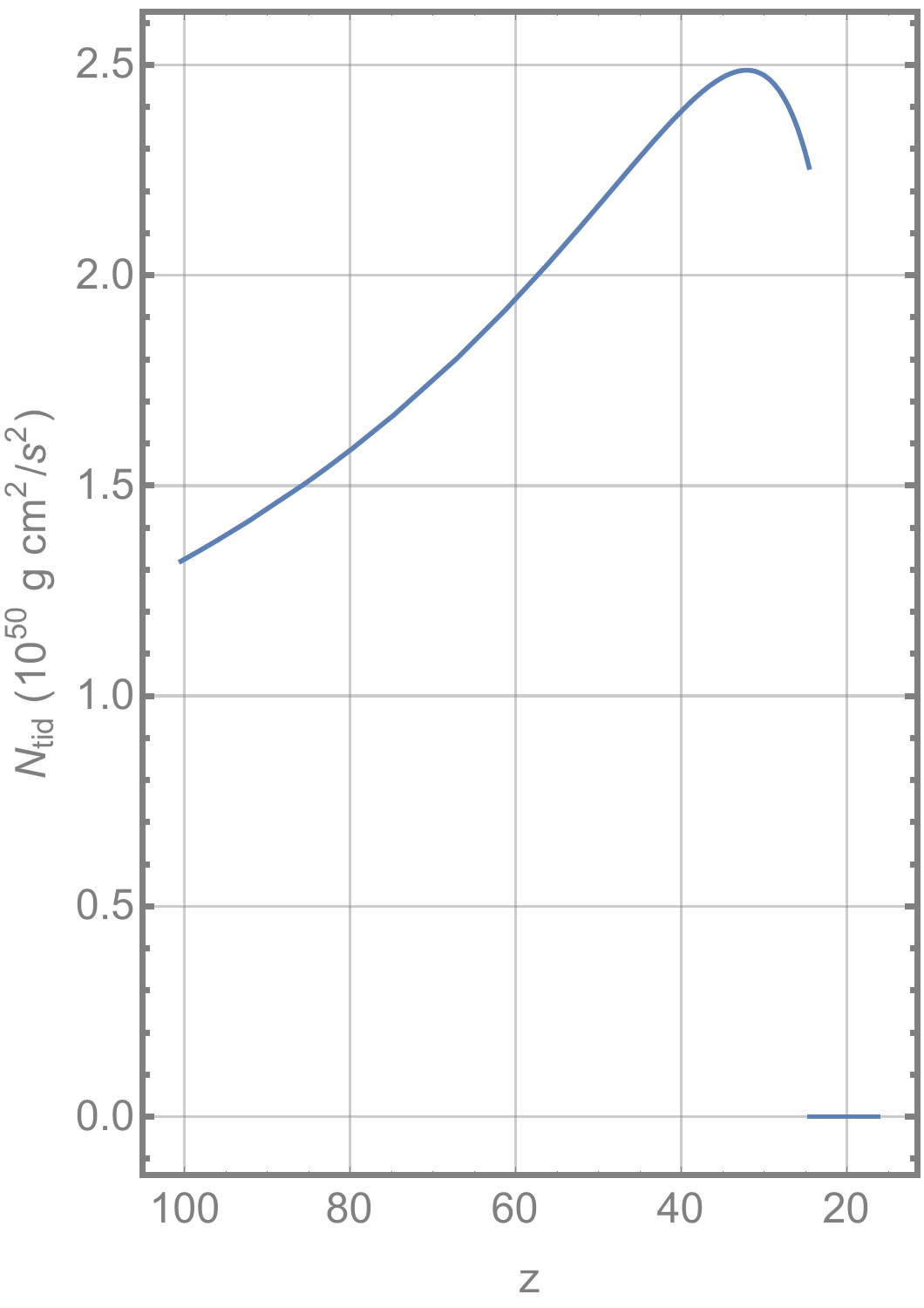}
\caption {Evolution, as a function of redshift $z$, of the cloud radius $r_{\rm cl}$ (top left panel), the cloud density $\rho_{\rm cl}$ in terms of background density (top right panel), the cloud moment of inertia $I_{\rm cl}$ (bottom left panel), and the tidal torque $N_{\rm tid}$ acting on the cloud (bottom right panel). Here the redshift of maximum radius $z_{\rm max}\approx24$ and that of virialization $z_{\rm vir}\approx16$, and $M_h=10^8 M_\odot$.}
\label{fig:halo_params}
\end{center}
\end{figure*}

Figure~\ref{fig:halo_params} show the evolution with redshift of the cloud radius $r_{\rm cl}(t)$ ($= r(t)$ given by Eq.~(\ref{eq:top-hat}) for $t<t_{\rm vir}$), the relative density of the cloud $\rho_{\rm cl}(t)/\rho_{\rm ext}(t)$, where $\rho_{\rm cl}$ is the density of the cloud, the moment of inertia of the cloud $I_{\rm cl}(t)$, and $N_{\rm tid}(t)$. Here $\rho_{\rm cl}(t)$ and $I_{\rm cl}(t)$ are defined as
        \begin{eqnarray}
            \rho_{\rm cl}(t) & = & \frac{3M_b}{4\pi r(t)^3} , \\
            I_{\rm cl}(t) & = & k M_t r(t)^2 ,
           \end{eqnarray}
where $M_t=M_h+M_b$, with $M_b$ being the cloud baryon/gas mass, and $k=2/5$ (where we have assumed the cloud to be a solid sphere). Figure~\ref{fig:halo_params} corresponds to $M_h=10^8 M_\odot$. For all our computations we use $\Omega_c h^2=0.12$, $\Omega_b h^2=0.022$, and $h=0.67$, where $H_0=100 h$ km/s/Mpc.

\subsection{Magnetic braking and the evolution of the angular speed}

A frozen-in magnetic field links the cloud with the external (background) medium of density $\rho_{\rm ext}$. To make this problem analytically  tractable we assume cylindrical symmetry with the $z$ axis being the axis of symmetry. We assume that the frozen-in magnetic field has only $r$ and $\phi$ components, with the field strength at the cloud surface $B_i=B(t_i) = B_0 \{1+z(t_i)\}^2$, where $B_0$ is the value of magnetic field strength at $z=0$. Also, even though the radius of the gas cloud is evolving, we choose a fixed radius $R = r_{\rm vir}$ to simplify the problem. Since the radius of the cloud does not change significantly until close to virialization of the halo (see the top left panel of Fig.~\ref{fig:halo_params}), this assumption does not significantly affect our main results. To render the problem analytically tractable, we also assume that the density ratio $\rho=\rho_{\rm cl}/\rho_{\rm ext}$ is constant\footnote{With the assumption of a constant $R$, this means that $\rho_{\rm cl}$ and $\rho_{\rm ext}$ are time independent.} and compute for three fixed values of $\rho$ = 1, 10, and 100, to cover the range of relative density that the evolving cloud achieves during its dynamical evolution until the point of virialization of its host halo (see the top right panel of Fig.~\ref{fig:halo_params}, which shows that $\rho$ varies from 1 to $\simeq 20$). Since our main interest is to compare the rotation speed of the cloud (at the time of virialization) with the limit of Eq.~(\ref{eq:vphi_limit2}), we do not have to follow the dynamical evolution of the cloud beyond the time of virialization.

Under the assumption of flux-freezing, following MP79, in our case the Alfv{\'e}n wave equations, analogous to Eqs.~(11) and (14) of MP79, are 
\begin{align}
    \frac{\partial^2 \Omega(r,t)}{\partial t^2} &= \frac{B_i^2}{4 \pi \rho_{\rm ext}} \frac{R^2}{r^2} \left[ \frac{1}{r}\frac{\partial}{\partial r} \left( r \frac{\partial \Omega}{\partial r} \right) \right], \ \ \ \ r > R , \\
    \frac{\partial^2 \Omega(t)}{\partial t^2} &= \frac{B_i^2}{\pi \rho_{\rm cl} R}  \frac{\partial \Omega(r,t)}{\partial r} \Big\vert_{r=R} + \frac{\partial \alpha_{\rm tid}(t)}{\partial t}, \ \ \ \ r = R.
\end{align}
Here $\Omega(r,t)$ is the angular velocity at radius $r$ at time $t$; in what follows we use $\Omega_{\rm cl} = \Omega(r=R,t)$ as the angular velocity of the cloud. The azimuthal component of the velocity,  $v_\phi(r,t) = r \Omega(r,t)$. Also $\alpha_{\rm tid}(t)=N_{\rm tid}(t)/I_{\rm cl}$ where $N_{\rm tid}(t)$ is the torque due to tidal interaction with the surrounding density field and $I_{\rm cl}$ is the moment of inertia of the gas cloud.  Using the transformation $\xi = (r/R)^2$ and $\tau=2t/(R/v_{A0})$, where $v_{A0}=B_i(4\pi \rho_{\rm ext})^{-1/2}$ is the Alfv{\'e}n speed, we get,
\begin{align}
    \frac{\partial^2 \Omega(\xi,\tau)}{\partial \tau^2} &= \frac{1}{\xi} \frac{\partial}{\partial \xi} \left( \xi \frac{\partial \Omega(\xi,\tau)}{\partial \xi} \right) , \ \ \ \ \xi > 1 , \label{eq:angvel}\\
    \frac{\partial^2 \Omega(\xi,\tau)}{\partial \tau^2} &= \frac{2}{\rho}  \frac{\partial \Omega(\xi,\tau)}{\partial \xi} + \frac{R}{2 v_{A0}} \frac{\partial \alpha_{\rm tid}(\tau)}{\partial \tau} , \ \ \ \ \xi = 1 , \label{eq:clangvel}
\end{align}
where $\rho = \rho_{\rm cl}/\rho_{\rm ext}$.

The boundary and initial conditions are
\begin{center}
$\Omega(\xi=\infty,\tau)=0$, \ \ \ \ $\Omega(\xi=1,\tau>0)=\Omega_{\rm cl}(\tau)$, \\ \vspace{0.5em}
$\Omega(\xi>1,\tau\leq0)=0, \ \ \ \
\frac{\partial\Omega(\xi,\tau)}{\partial \tau}=0$ 
for $\xi>1$ and $\tau\leq 0$, \\ \vspace{0.5em}
$\Omega_{\rm cl} = 0$, for $\tau\leq 0$, \\ \vspace{0.5em}
$\frac{\partial\Omega_{\rm cl}}{\partial \tau}=0$ for $\tau<0$, $\frac{\partial\Omega_{\rm cl}}{\partial \tau}=\alpha_{\rm tid}(\tau=0)$ for $\tau = 0$.
\end{center}
We note that $\Omega_{\rm cl}=0$ for $\tau \leq 0$ and 
$\partial\Omega_{\rm cl}/\partial\tau (\tau=0) =\alpha_{\rm tid}(\tau=0)$ 
in our case, while in MP79 the corresponding initial conditions are  $\Omega_{\rm cl}=0$ for $\tau < 0$ and $\Omega_{\rm cl}=\Omega_0$ for $\tau = 0$. We also assume that for $\tau<0$ the  $\phi$ component of the magnetic field $B_\phi=0$. The main difference between our case and that studied by MP79 is that we have a source that spins up the cloud (tidal interaction), a source that is switched on at high redshift in the expanding universe. 

As in the case of the MP79 equations, our Alfv{\'e}n wave equations can be solved by using the Laplace transform defined through 
\begin{align}
    {\tilde\Omega(\xi,s)} &= \int^\infty_0 d\tau\, e^{-s\tau} \Omega(\xi,\tau), \\
    {\Omega(\xi,\tau)} &= \int^{a+i\infty}_{a-i\infty}\!\!\!\! ds\, e^{s\tau} \tilde\Omega(\xi,s), \label{eq:laplacetransform}
\end{align}
where $a$ is a real number. Laplace transforming Eqs.~(\ref{eq:angvel}) and~(\ref{eq:clangvel}), we obtain
\begin{align}
s^2 {\tilde\Omega(\xi,s)}\ & - s\Omega(\xi,0) - \frac{\partial\Omega(\xi,\tau)}{\partial\tau}\Big\vert_{\tau=0} \nonumber \\ 
& - \frac{1}{\xi} \frac{\partial \tilde\Omega(\xi,s)}{\partial \xi} - \frac{\partial^2 \tilde\Omega(\xi,s)}{\partial \xi^2} = 0 , \ \ \ \ \xi > 1 , \label{eq:lapengvel} \\
s^2 {\tilde\Omega(\xi,s)}\ & - s\Omega(1,0) - \frac{\partial\Omega(1,\tau)}{\partial\tau}\Big\vert_{\tau=0} - \frac{2}{\rho} \frac{\partial \tilde\Omega(\xi,s)}{\partial \xi} \nonumber \\ 
& - \frac{R}{2 v_{A0}} \left( s\tilde\alpha_{\rm tid}(s) -\alpha_{\rm tid}(0)\right) = 0 , \ \ \ \ \xi = 1 , \label{eq:lapclangvel}
\end{align}
where
\begin{align}
{\tilde\alpha_{\rm tid}(s)} = \int^\infty_0 d\tau\, e^{-s\tau} \alpha_{\rm tid}(\tau) .
\label{eq:alphadef}
\end{align}

Substituting $\Omega(\xi,0)=0$, 
$\partial\Omega/\partial\tau(\xi>1,\tau)\vert_{\tau=0}=0$, 
and $\partial\Omega/\partial\tau(\xi=1,\tau)\vert_{\tau=0}=0$\footnote{This is an approximation to the initial condition. More correctly it should be $\partial\Omega/\partial\tau(\xi=1,\tau)\vert_{\tau=0} = \alpha_{\rm tid}(\tau=0) = \alpha_{\rm tid}(\theta_i)$ where $\theta_i=\cos^{-1}\left((1-\delta_i)/(1+\delta_i)\right)$. The approximation we use here has an insignificant effect on our results.} 
into Eqs.~(\ref{eq:lapengvel}) and (\ref{eq:lapclangvel}), we find
\begin{align}
    s^2 {\tilde\Omega(\xi,s)} -\frac{1}{\xi} \frac{\partial \tilde\Omega(\xi,s)}{\partial \xi} - \frac{\partial^2 \tilde\Omega(\xi,s)}{\partial \xi^2} &= 0 , \ \ \ \ \xi > 1, \label{eq:lapsol1}\\
    s^2 {\tilde\Omega(\xi,s)} - \frac{2}{\rho} \frac{\partial \tilde\Omega(\xi,s)}{\partial \xi} - \frac{sR}{2 v_{A0}} \tilde\alpha_{\rm tid}(s) &= 0 , \ \ \ \ \xi = 1. \label{eq:lapsol2}
\end{align}

\begin{figure}
\includegraphics[width=0.49\textwidth]{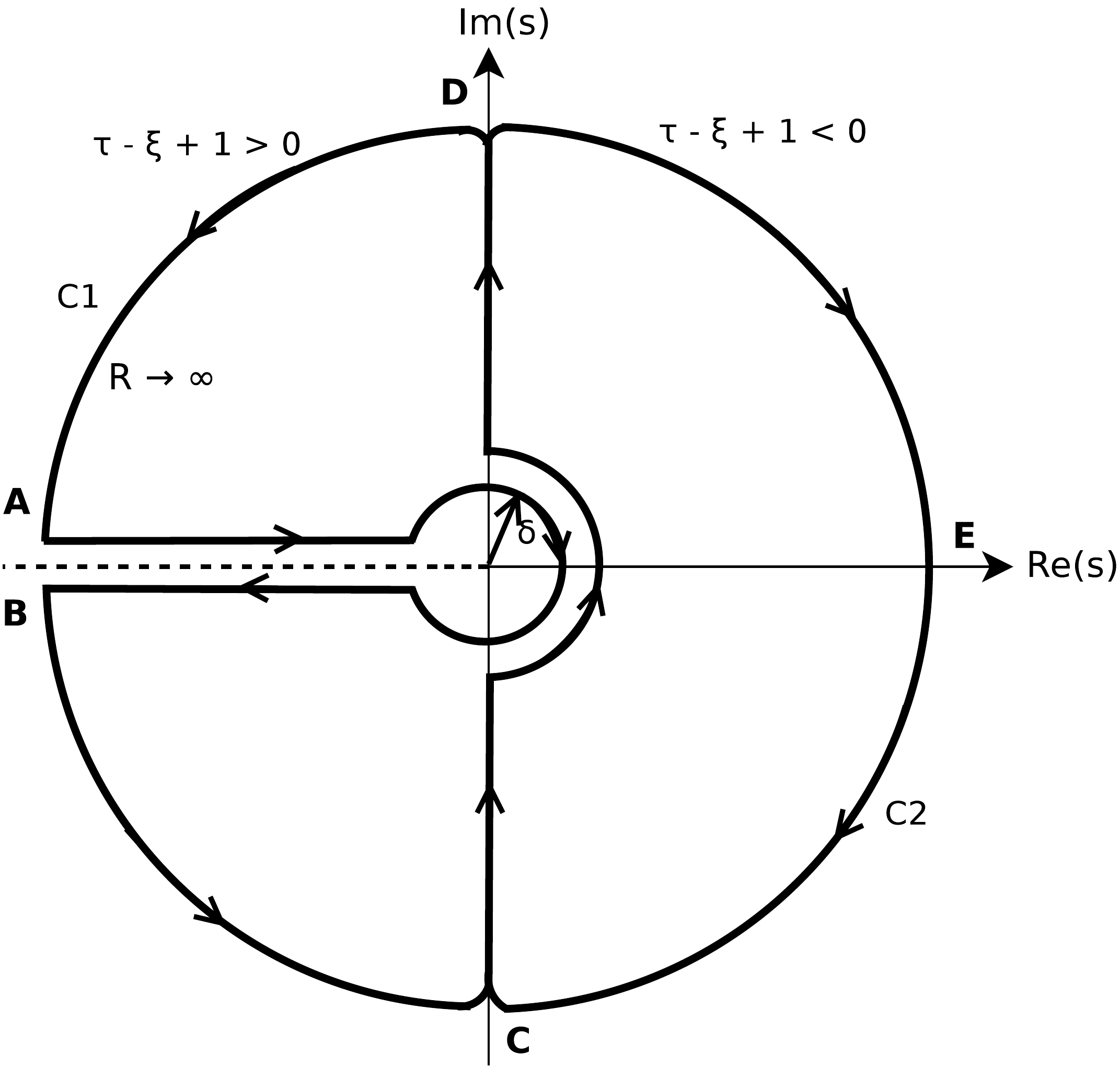}
\caption {The MP79 contours for doing the integral in Eq.~(\ref{eq:lapsolfin}). The contour is closed  on the right of the imaginary $s$ axis, CDEC (C2) or on the left of the axis, CDABC (C1), depending on the value of $\tau$.}
\label{fig:contour}
\end{figure}

Equation~(\ref{eq:lapsol1}) has the solution (see MP79, Eqs.~(24a) and (25)) 
\begin{align}
{\tilde\Omega(\xi,s)} &= F(s) K_0(\xi s) ,
\label{eq:lapsol3} 
\end{align}
where $K_0$ is a modified Bessel function of the second kind, and we determine $F(s)$ by substituting Eq.~(\ref{eq:lapsol3}) into Eq.~(\ref{eq:lapsol2}) to get
\begin{align}
F(s) &= \Omega_0(s) \left[ s K_0(s) - \frac{2}{\rho} K'_0(s) \right]^{-1} , 
\label{eq:lapsol25}
\end{align}
where we have defined $\Omega_0(s) = (R/2 v_{A0}) \tilde\alpha_{\rm tid}(s)$.

Substituting Eqs.~(\ref{eq:lapsol3}) and~(\ref{eq:lapsol25}) into Eq.~(\ref{eq:laplacetransform}) gives
\begin{align}
{\Omega(\xi,\tau)} &= \frac{1}{2 \pi i} \int^{a+i\infty}_{a-i\infty} \!\!\!\! ds\, e^{s\tau} \frac{{\Omega_0(s)} K_0(\xi s)}{s K_0(s) - \frac{2}{\rho} K'_0(s)} , \ \ \ \ \xi \geq1.
\label{eq:lapsolfin}
\end{align}
Equation~(\ref{eq:lapsolfin}) can be solved by integration in the complex plane. Suitable contours are  shown in Fig.~\ref{fig:contour}. The contours are  the same as those used in MP79 as the integrand in our case is  $\eta(s) = \Omega_0(s)\eta_M(s)$,  where the integrand in MP79 is $\eta_M(s) = {\exp(s\tau) K_0(\xi s)}/{f(s)}$ with $f(s)=sK_0(s) + \lambda K_1(s)$ where $K_1=-K'_0$ and $\lambda = 2/\rho$. The additional factor $\Omega_0(s)$ here doesn't introduce any new singularities. The integrand  $\eta(s)$ has two simple poles at $s_1$ and $s_2$ where $s_{1,2} = -\alpha \pm i\beta$ (Re($s_{1,2}) < 0$) are the zeros of $f(s)$. The integrand has a branch point at $s= 0$  which can be avoided by a branch cut along the negative real axis and a small loop around $s=0$. 

To evaluate the integral in Eq.~(\ref{eq:lapsolfin}) we can close the Fig.~\ref{fig:contour} contour either to the right of the imaginary $s$ axis (CDEC or C2) or to the left of the axis (CDABC or C1). In terms of contour integrals, Eq.~(\ref{eq:lapsolfin}) is
\begin{align}
& 2\pi i \ \Omega(\xi,\tau) = \int_{CD} \!\! \eta(s) ds = \int_{C2} \!\! \eta(s) ds - \int_{DEC} \!\! \eta(s) ds \label{eq:right-contr} \\
& = \int_{C1} \!\! \eta(s) ds - \int_{DA} \!\! \eta(s) ds - \int_{BC} \!\! \eta(s) ds - \int_{AB} \!\! \eta(s) ds . \label{eq:left-cntr}
\end{align}

The integral along the tiny loop ($\delta\rightarrow 0$) around $s =0$ is zero (MP79).\footnote{We have $\eta(s)=\Omega_0(s)\eta_M(s) = \tau_0 \tilde\alpha_{\rm tid}(s)\eta_M(s);$ since $ \eta_M(s)\to0$ when $s\to0$ it follows that $\eta(s)\to0$ when $s\to0$} Now by employing the asymptotic form of $\eta_M(s)$ it can be shown that
\begin{align}
\lim_{s\to\infty}\eta_M(s) &= 0 \ \ \ \ {\textrm{ if $\xi > 1+ \tau$}} , \\
\lim_{s\to-\infty}\eta_M(s) &= 0 \ \ \ \ {\textrm{ if $\xi < 1+ \tau$}} , 
\end{align}
and at the cloud radius, $\xi=1$,
\begin{align}
\label{eq:31}
\lim_{s\to\infty}\eta_M(s) &= 0 \ \ \ \ {\textrm{ if $\tau < 0$}} , \\
\lim_{s\to-\infty}\eta_M(s) &= 0 \ \ \ \ {\textrm{ if $\tau > 0$}} .
\end{align}
These indicate that to find the evolution of the angular velocity for $\tau>0$ we must use the contour C1 for the integration, as the integration along the contour C2 does not converge for $\tau>0$. Also we find $\Omega_{\rm cl}(\tau) = 0$ for $\tau < 0$,  since Eq.~(\ref{eq:31}), the fact that contour C2 does not contain any poles, and $\Omega_0(s) \to 0$ as $s\to\infty$ imply that the right hand side of Eq.~(\ref{eq:right-contr}) vanishes.

The contour integral on the left side of the imaginary $s$ axis (C1) is more complicated as it involves computing residues for two simple poles and the integration along $AB$.  First, the residue for the two simple poles can be readily computed
\begin{align}
& \sum_{n=1}^2 {\rm Res} \ \eta(s_n) = \nonumber \\
& \ \ \frac{{\Omega_0(s_1)} \exp(s_1\tau) K_0(\xi s_1)}{f'(s_1)} + \frac{{\Omega_0(s_2)} \exp(s_2\tau) K_0(\xi s_2)}{f'(s_2)},
\label{eq:residue}
\end{align}
where $f'(s) = \left( 2-\lambda + s^2/\lambda\right)K_0(s)$. However, in our case, unlike for MP79, 
\begin{align}
\lim_{s\to-\infty}\eta(s) &\neq 0 \ \ \ \ {\textrm{ as }} \lim_{s\to-\infty}\Omega_0(s)=\infty .
\end{align}
That is, the integrand does not converge for all $\tau>0$  because, for $s\to-\infty, \tilde\alpha_{\rm tid}(s)\to\infty$, Eq.~(\ref{eq:alphadef}).
This situation can be avoided if we assume $\alpha(\tau)\to 0$ for some $\tau > \tau_{\rm max}$. We assume $\alpha(t)\to 0$ for $t>t_{\rm max}$ where $t_{\rm max}$ is the time at which the cloud is at its maximum radius and $\tau_{\rm max} = \tau (t_{\rm max})$. This assumption is consistent with the analysis of \citet{1988ApJ...329..589R}. This allows us to establish that\footnote{Since $\tilde\alpha_{\rm tid}(s)=\int_0^\infty \alpha(\tau) e^{-s\tau}d\tau \equiv \int_0^{\tau_{\rm m}} \alpha(\tau) e^{-s\tau}d\tau < \int_0^{\tau_{\rm m}} \alpha_{\rm m} e^{-s\tau}d\tau = \alpha_{\rm m}(1-\exp(-s\tau_{\rm m})/s)$ which implies $\Omega_0(s)=(\tau_0 \alpha_{\rm m}/s)\{1-\exp(-s\tau_{\rm m})\}$ so $\eta(s)=(\tau_0\alpha_{\rm m}/s)\{\exp(s\tau)-\exp[s(\tau-\tau_{\rm m})]\}/(s+\lambda)$ where $\tau_{\rm m}$ is $\tau(t_{\rm max})$ and $\alpha_{\rm m}$ is the maximum value of $\alpha(\tau)$ in the interval $0<\tau<\tau_{\rm m}$.}
\begin{align}
\lim_{s\to-\infty}\eta(s) &= 0 \ \ \ \ {\textrm{ if $\tau > \tau_{\rm max}$  }}. 
\end{align}

The evolution of the rotation of the cloud for the time range $\tau>\tau_{\rm max}$ can now be written as, using Eq.~(\ref{eq:left-cntr}), 
\begin{align}
\Omega_{\rm cl}(\tau> \tau_{\rm max}) & = - \frac{1}{2\pi i} \int_{AB} \eta(s) ds + \sum_{n = 1}^{2} {\rm Res} \ \eta(s_n) .
\end{align}
As noted above,  it can be shown that the integral along the tiny loop ($\delta\rightarrow 0$) around the origin is zero. Using the transformation $s=x\exp(i\theta)$ and Eq.~(\ref{eq:residue}) the above equation can be rewritten as
\begin{align}
& \Omega_{\rm cl}(\tau> \tau_{\rm max})  = {\mathrm Re}\left[ \sum_{n=1}^{2} \frac{{\Omega_0(s_n)} \exp(s_n\tau) K_0(s_n)}{\left( 2-\lambda + s_n^2/\lambda\right)K_0(s_n)} \right] \nonumber \\ 
& \ \ \ \ - \frac{1}{2\pi i} \int_\infty^{\delta\to0} d[x e^{i\pi}]\eta[x e^{i\pi}] -\frac{1}{2\pi i} \int^\infty_{\delta\to0} d[x e^{-i\pi}]\eta[x e^{-i\pi}] \nonumber \\ 
& = {\mathrm Re}\left[ \sum_{n=1}^{2} \frac{{\Omega_0(s_n)} \exp(s_n\tau) K_0(s_n)}{\left( 2-\lambda + s_n^2/\lambda\right)K_0(s_n)} \right] \nonumber \\
& \ \ \ \ - \frac{1}{2\pi i} \int_0^\infty dx \left( \eta[x e^{i\pi}] - \eta[x e^ {-i\pi}] \right) .
\label{eq:2ndlast}
\end{align}
Using $K_\nu[x \exp(\pm i\pi)] = \exp(\mp i\pi\nu) K_\nu(x) \mp i\pi I_\nu(x)$ (\citet{Watson1952} p. 152), where $I_\nu$ is the modified Bessel function of the first kind, we find, after some simplification, 
\begin{align}
& \eta[x e^{i\pi}] - \eta[x e^{-i\pi}] = \nonumber \\ 
& \ \ \ \ \frac{2\pi i \lambda \exp(-x\tau) \Omega_0(-x)}{x\left[ xK_0(x)+\lambda K_1(x) \right]^2 + \pi^2\left[ xI_0(x) - \lambda I_1(x) \right]^2}, 
\label{eq:branchInt}
\end{align}
which when used in Eq.~(\ref{eq:2ndlast}), gives the final expression for the angular velocity of the cloud
\begin{align}
& {\Omega_{\rm cl}(\tau> \tau_{\rm max})} = {\mathrm Re}\left[ \sum_{n=1}^{2} \frac{{\Omega_0(s_n)} \exp(s_n\tau) K_0(s_n)}{\left( 2-\lambda + s_n^2/\lambda\right)K_0(s_n)} \right] \nonumber \\
& - \lambda \int_0^\infty d ({\rm ln}x) \frac {{\Omega_0(-x)} \exp(-x\tau)} {\left[ xK_0(x)+\lambda K_1(x) \right]^2 + \pi^2\left[ xI_0(x) - \lambda I_1(x) \right]^2} . 
\label{eq:omegacl}
\end{align}

\begin{figure*}
\begin{center}
\includegraphics[width=0.445\textwidth]{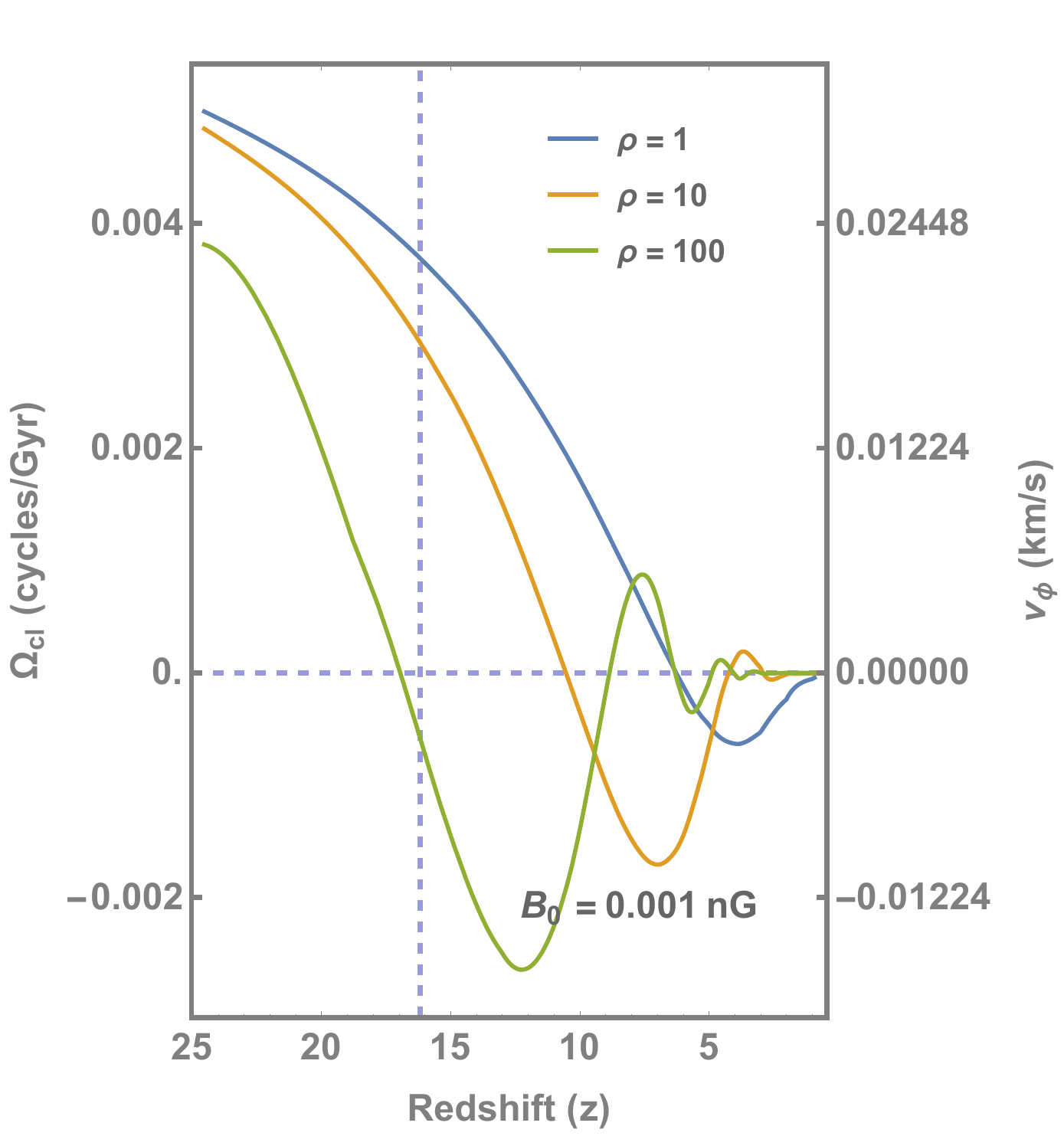}
\hspace{2em}
\includegraphics[width=0.460\textwidth]{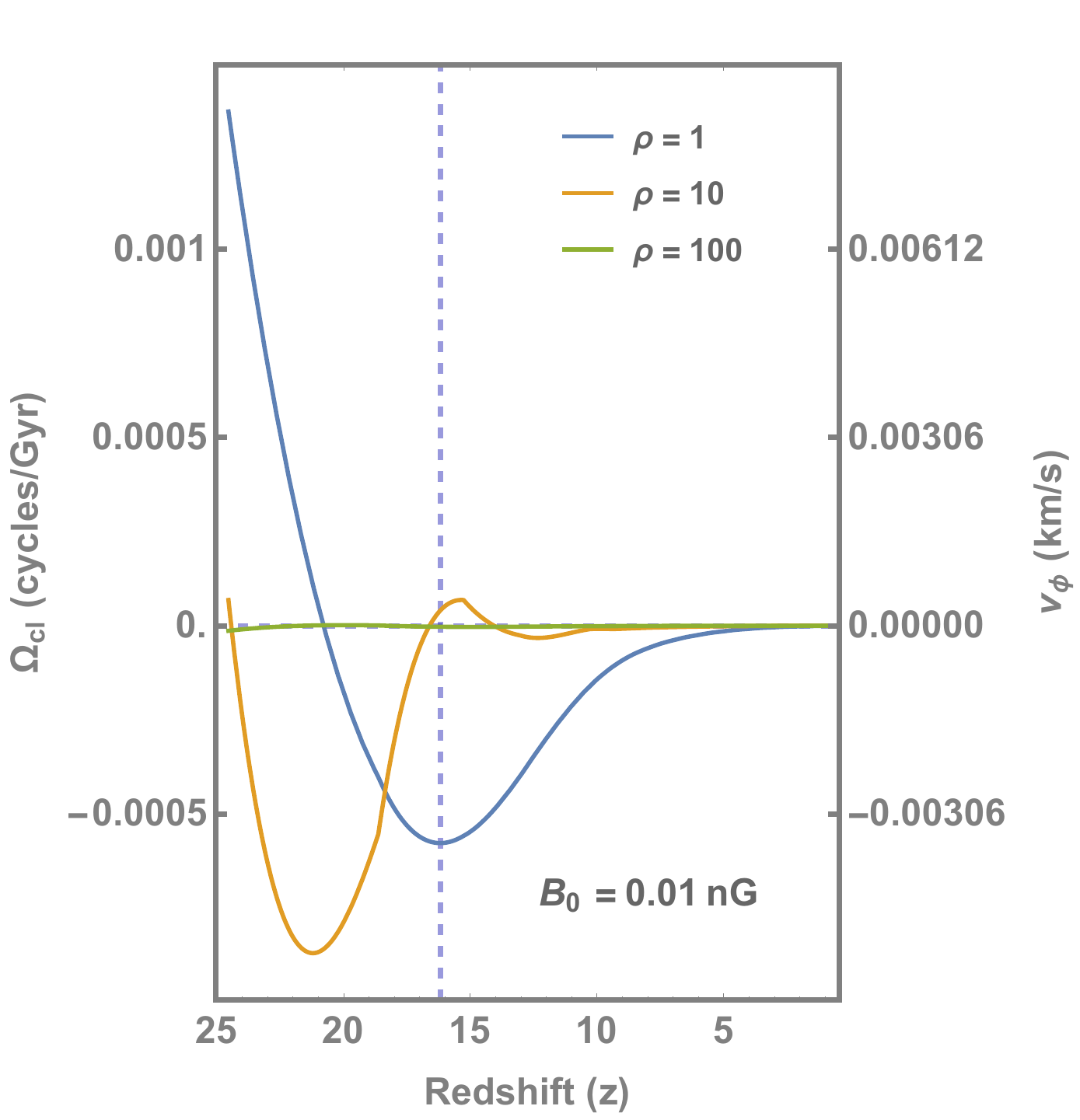}
\caption {Evolution of rotation speed $v_\phi$ (or $\Omega_{\rm cl}$) for halo mass $M_h= 10^8 {\rm M}_\odot$. See main text for description and discussion.}
\label{fig:linear_plots}
\end{center}
\end{figure*}

\begin{figure*}
\begin{center}
\includegraphics[width=1.0\textwidth]{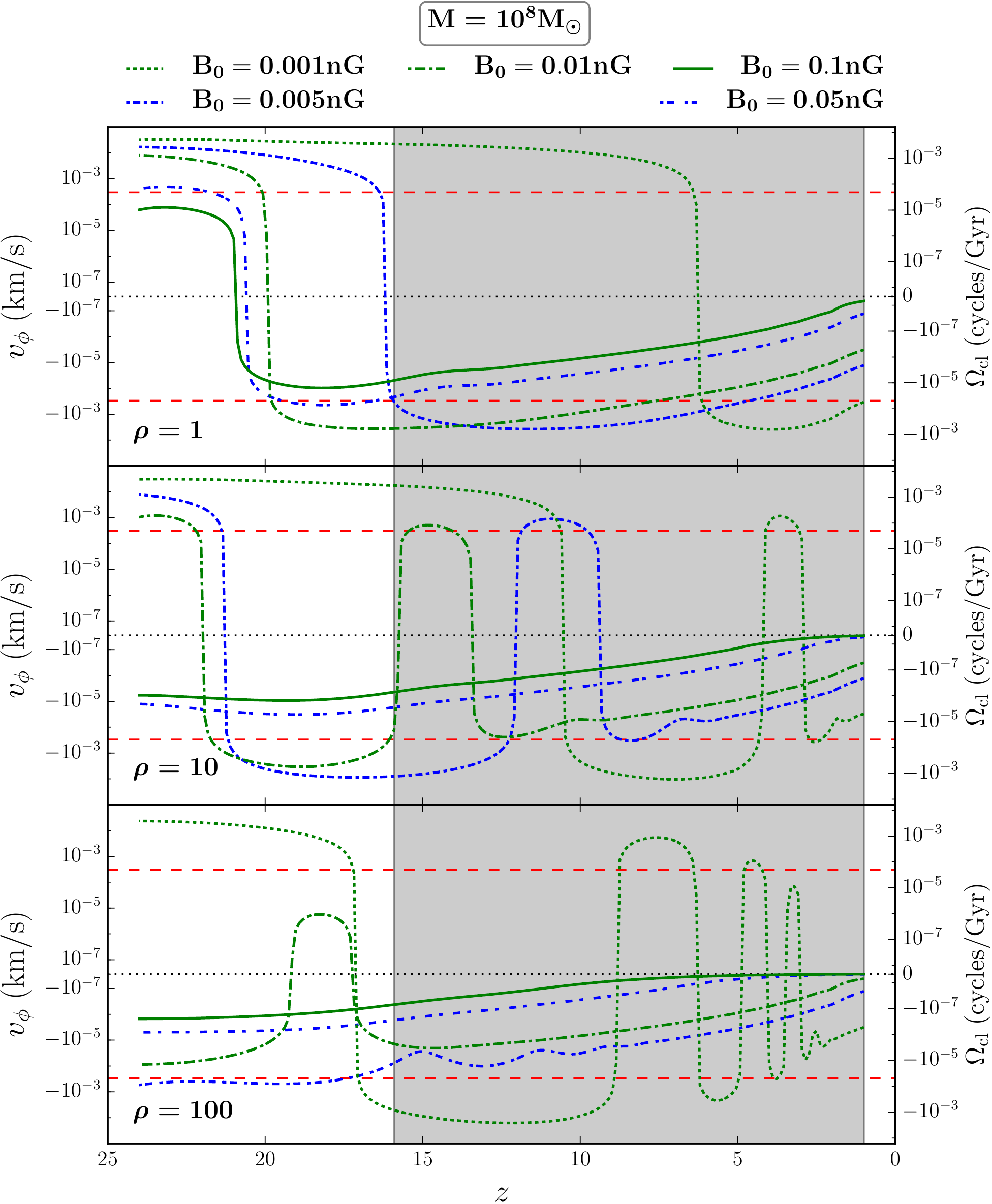}
\caption {Evolution of $v_\phi$ (or $\Omega_{\rm cl}$) for halo mass $M_h= 10^8 {\rm M}_\odot$. See main text for description and discussion.}
\label{fig:mh_10_8}
\end{center}
\end{figure*}

\begin{figure*}
\begin{center}
\includegraphics[width=1.0\textwidth]{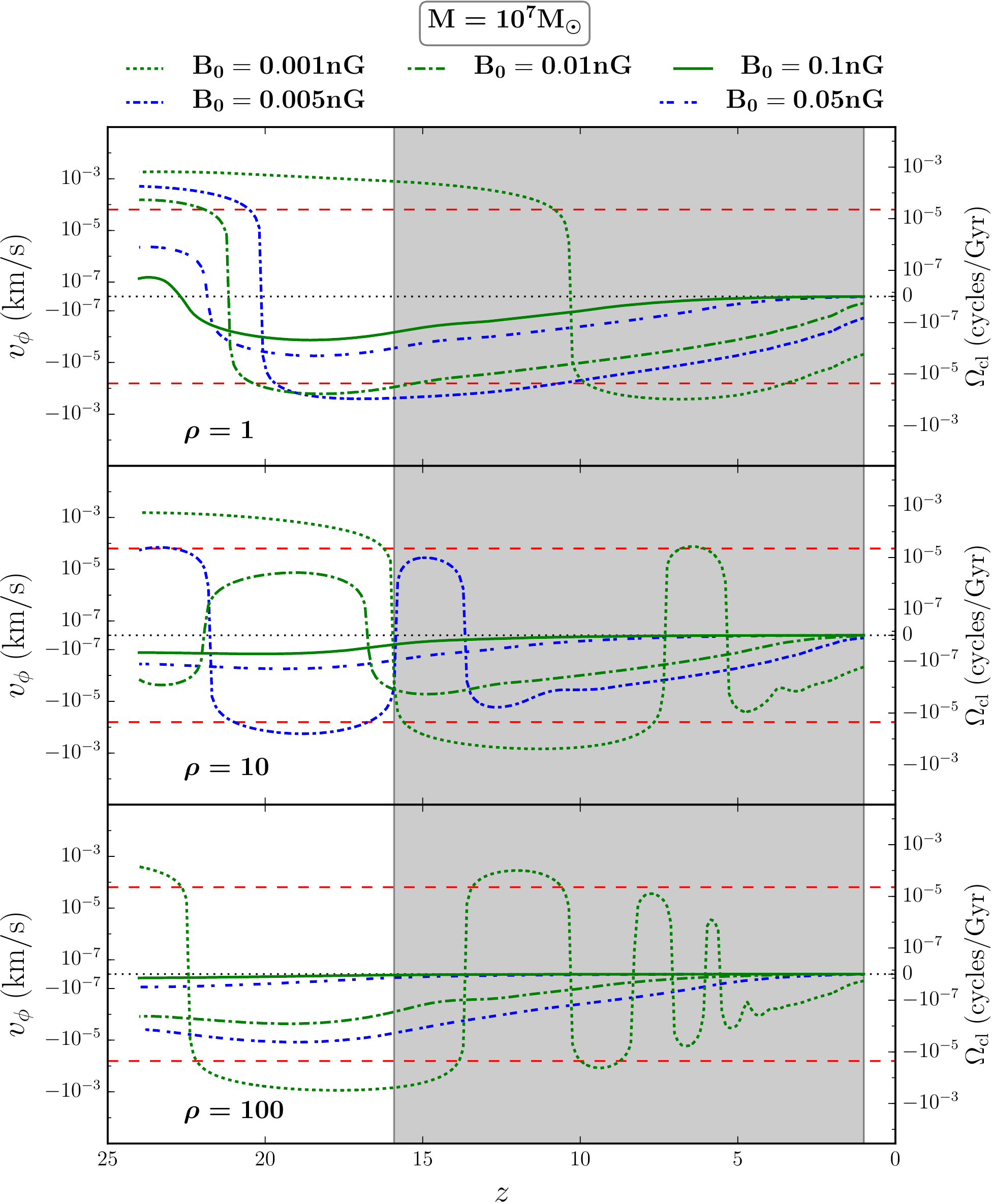}
\caption {Evolution of $v_\phi$ (or $\Omega_{\rm cl}$) for halo mass $M_h= 10^7 {\rm M}_\odot$. See main text for description and discussion.}
\label{fig:mh_10_7}
\end{center}
\end{figure*}

\begin{figure*}
\begin{center}
\includegraphics[width=1.0\textwidth]{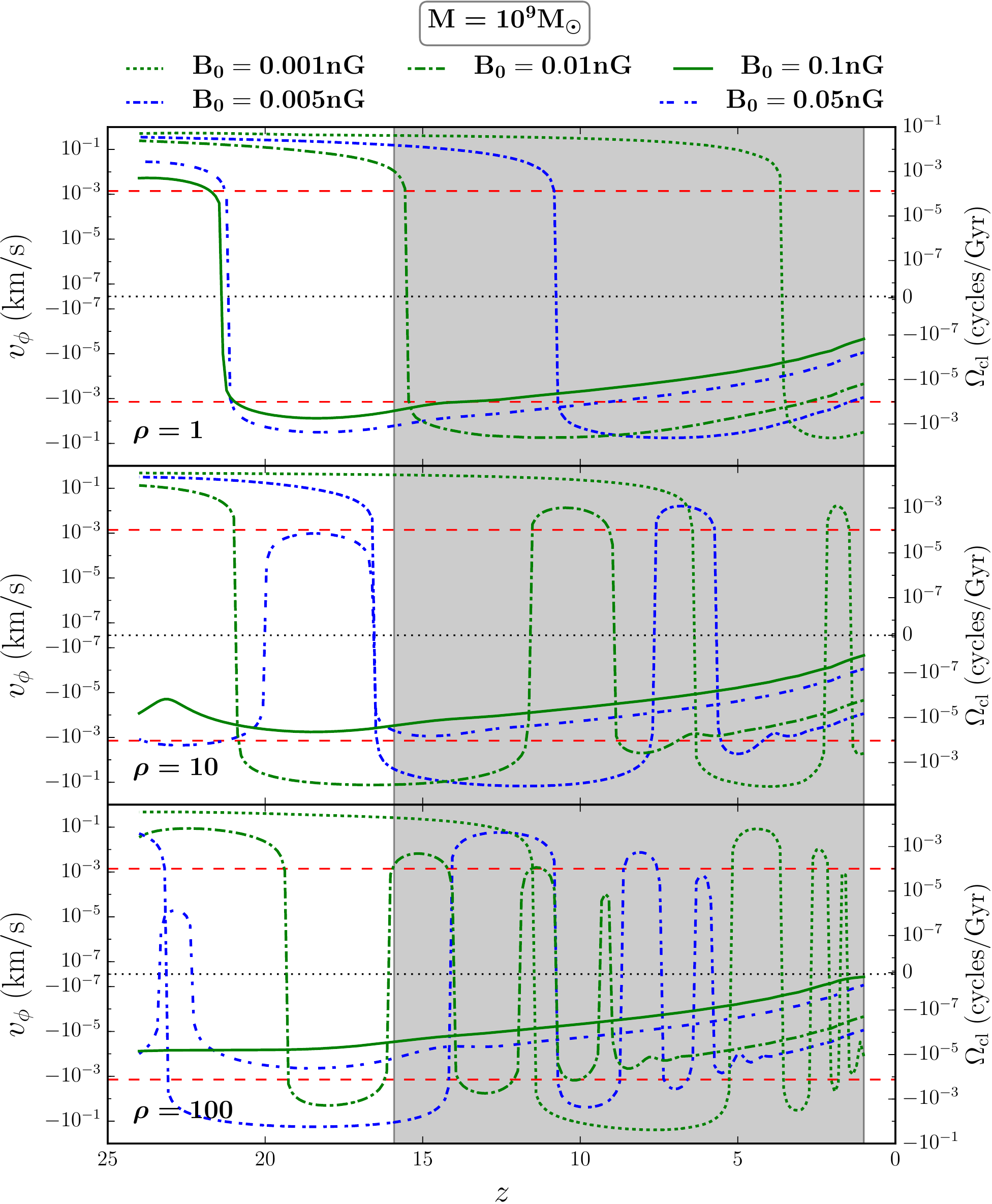}
\caption {Evolution of $v_\phi$ (or $\Omega_{\rm cl}$) for halo mass $M_h= 10^9 {\rm M}_\odot$. See main text for description and discussion.}
\label{fig:mh_10_9}
\end{center}
\end{figure*}

\section{Results}

In our framework, both cloud rotation  and magnetic braking vanish at the initial time: $v_\phi=0$,  $B_\phi=0$ at  $\tau=0$. As soon as there is some build up of angular velocity due to tidal torque, $B_\phi$ also starts  increasing, enhancing  the magnetic braking. If the time scale of  magnetic braking is  short in comparison to the angular velocity build up time scale due to the tidal torque,  the angular velocity will be driven towards zero on the magnetic breaking time scale.  However,  as the angular velocity approaches  zero, the  magnetic braking  also become weaker and tidal torque causes the angular momentum to build up again. This cycle continues as long as the tidal torque remains on, and the solution $v_\phi$ is  oscillatory during this phase with its amplitude determined by the relative time scales of tidal torquing and magnetic braking. As soon as the tidal torque becomes zero magnetic braking takes over and reduces the angular momentum to zero on a  time scale which depends on the magnetic field strength. 

Figure~\ref{fig:linear_plots} show the evolution of $\Omega_{\rm cl}$, for the time range $\tau>\tau_{\rm rmax}$,  Eq.~(\ref{eq:omegacl}), for different values of $\rho=\rho_{\rm cl}/\rho_{\rm ext}$ and $B_0$; $t_{\rm rmax}$ corresponds to the time at which the halo reaches its maximum radius and the tidal torque is switched off (Ryden 1988). The vertical blue-dashed lines in the two panels indicate $z_{\rm vir}$ below which our assumption of constant density becomes progressively more inaccurate. Figure~\ref{fig:linear_plots} shows that the angular velocity  behaves like a damped oscillator during this time period. 

For a better representation of our results we have plotted $v_\phi$ for different cases on plots that have a symmetrical-logarithmic $y$ axis scale, in Figs.~\ref{fig:mh_10_8}, \ref{fig:mh_10_7}, and \ref{fig:mh_10_9} respectively for the cases $M_h = 10^8, 10^7$, and $10^9 M_\odot$. In these figures the red-dashed lines represent the corresponding critical $v_\phi^{\rm crit}(t_{\rm vir})$ given by Eq.~(\ref{eq:vphi_limit2}). The region past the time of virialization of the halo is shaded because our assumption of constant density becomes progressively worse for these times. This however does not affect the main result, which is that the rotation speed at the time of virialization $v_\phi(t_{\rm vir})$ is smaller than $v_\phi^{\rm crit}(t_{\rm vir})$, and continues to decrease. For example, let us examine the case $M_h = 10^8 M_\odot$ (Fig.~\ref{fig:mh_10_8}). Here we see that for the density ratio $\rho=1$ as the magnetic field strength $B_0$ approaches $\sim 0.1$~nG magnetic braking reduces the rotation speed $v_\phi$ to $\lesssim 10^{-4}$ km s$^{-1}$ by the time of virialization ($z_{\rm vir}=16$).  Also notice that $v_\phi$ never goes higher than $v_\phi^{\rm crit}$. In this case we can safely conclude that for  magnetic field strength $B_0 \gtrsim 0.1$ nG a collapsing cloud of total mass  $10^8 M_\odot$  can lose sufficient  initial angular momentum due to magnetic braking. Importantly, the figures show  that  as the density ratio $\rho$ increases, which would be the case for a collapsing gas cloud, magnetic braking becomes even more efficient ($v_\phi \lesssim 10^{-6}$ km s$^{-1}$ for $\rho=100$). 

These figures also show that the magnetic field strength needed to remove the initial angular momentum scales inversely with the mass of the halo. For $M_h = 10^7 \rm M_\odot$,  magnetic field $B_0$ $\sim 0.01$ nG (for $\rho=1$) is sufficient to remove the angular momentum whereas for $M_h=10^9 M_\odot$, the required field strength is $B_0$ $\gtrsim 0.1$~nG  for $\rho=10$.  The critical rotation speed also  scales inversely with the halo mass, Eq.~(\ref{eq:vphi_limit2}), but we notice that  larger magnetic fields are needed to remove angular momenta for more massive halos, even though the dependence on the halo mass is weak.

The current most stringent bound on a primordial magnetic field strength is $B_0 \lesssim 0.1\hbox{--}0.6$~nG  at 1~Mpc (comoving) scale for $n_{\rm B} \sim -2.9$  \citep{2016RPPh...79g6901S,2016A&A...594A..19P,2015MNRAS.451.2244C,2015MNRAS.451.1692P,2013ApJ...762...15P,2012ApJ...748...27P,2012PhRvL.108w1301T}.\footnote{The scale invariant magnetic field power spectral index is $n_{\rm B} = -3$, which is a divergent case. Close to this limit inflation generates a large enough cosmological magnetic field \citep{1992ApJ...391L...1R} to explain galactic magnetic fields in the anisotropic collapse and differential rotation amplification scenario \citep{1970AuJPh..23..731P,1988plap.work..197K} and also generates a close to scale-invariant power spectrum of spatial inhomogeneity density perturbations consistent with what is observed in the cosmic microwave background anisotropy.}  
A cloud of mass $M_h = 10^8 M_\odot$ reaches its maximum radius of $\simeq 1$~kpc (at $z_{\rm max}\simeq24$) which corresponds roughly  to a comoving wavenumber $k \simeq 0.04 \ {\rm kpc}^{-1}$. The corresponding RMS magnetic field smoothed over a scale $k$ is $\bar{B}(k) = B_0 (k/k_0)^{(n_{\rm B}+3)/2}$ \citep{2005MNRAS.356..778S,2011ApJ...726...78K}. For a near scale-invariant spectral index $n_{\rm B} = -2.9$ a primordial magnetic field of strength $B_0 = B_{z=0} = 0.1$ nG at 0.04 kpc$^{-1}$  corresponds to a magnetic field strength of $\simeq 0.12$~nG  smoothed at 1~Mpc. This is comparable to the magnetic field needed to remove angular momentum for a $10^9 \rm  M_\odot$ halo. In other words, current primordial magnetic field bounds are consistent with the hypothesis that a primordial magnetic field could play an important role in transporting away the initial angular momentum from collapsing gas on mass scales relevant to the formation of SMBHs.

\section{Discussion and conclusions}

The main challenge for the DCBH model of SMBH formation is to find a way to efficiently dispose of angular momentum \citep[see, e.g.][]{2018ApJ...853L...3H,2018MNRAS.478.3961S}. Various ways of doing this have been proposed, such as radiation drag against the cosmic microwave background at very high redshift \citep{1993ApJ...419..459U,2018JApA...39....9P}, viscosity driven by magnetic fields or turbulence \citep{2003ApJ...598L...7C}, and self-gravitational instabilities  due to turbulent flow \citep{1969MNRAS.144..425P,1997ApJ...480..167M,2009ApJ...702L...5B}. Another class of possibilities include forming the DCBH out of low angular momentum material, either in halos with very low angular momenta \citep{1995ApJ...443...11E} or in the low angular momentum tail of material in halos \citep{2004MNRAS.354..292K}. The latter scenario requires the removal of a substantial amount of angular momentum from the in-falling gas at a very early stage for it to be able to form a central massive object which ultimately collapses into a DCBH as a result of post-Newtonian gravitational instabilities \citep{2010RPPh...73a4901L}. In this work we study and strengthen the \citet{1995Princeton} proposal that the presence of a sufficiently strong primordial cosmological magnetic field could provide enough magnetic braking to remove the angular momentum from in-falling material at the early stages of the collapse.

Many studies related to  possible effects of magnetic fields on the formation of early massive stars and black holes have  found that an initial weak magnetic field  could  amplify  considerably during gravitational collapse and become dynamically relevant, which could play a role in  suppressing fragmentation \citep{2016A&A...585A.151L,2014MNRAS.440.1551L,2013A&A...553L...9V,2010ApJ...721..615S,2009ApJ...703.1096S}.
The idea of magnetic braking due to frozen-in magnetic fields has been extensively used in the context of star formation scenarios of normal population stars to early Pop-III and supermassive stars, \citep{2019PhRvD..99f4057S,2013MNRAS.435.3283M,2012ApJ...754L..26M,2011A&A...525L..11M,1995ApJ...452..386B,1994ApJ...432..720B}. From timescale estimates \citet{1995Princeton} showed that magnetic braking by a cosmological magnetic field of strength needed to explain galactic magnetic fields in the anisotropic collapse and differential rotation amplification scenario \citep{1970AuJPh..23..731P,1988plap.work..197K} would be able to remove angular momentum during the cosmological formation of primordial Pop III stars and black holes (in the DCBH formation picture). In this paper we present the first study of the dynamical effect of magnetic braking in the context of DCBH formation at high redshifts. To do this we adapted the formalism developed in MP79 to allow for the  simultaneous buildup of angular momentum in halos at early times due to tidal interactions with the surrounding  inhomogeneous density field. 

Primordial magnetic fields are modeled as Gaussian random with zero mean and a given RMS. In our study, we show that magnetic field strength of greater than  0.1 nG is needed for magnetic braking. If this strength corresponds to RMS then nearly 30\% of objects in the mass range of interest could undergo magnetic braking and the SMBH formation could become too frequent an event. However, depending on how many sigma fluctuation of the magnetic field is 0.1~nG, a smaller fraction of objects would undergo magnetic braking. For instance, if the RMS of the magnetic field is 0.05~nG then less than 10\% of objects would be affected by magnetic braking. Therefore, magnetic braking could be a frequent  or a rare  event depending on the RMS of the magnetic field (e.g. \cite{2010ApJ...721..615S}). As the number of BH precursors  needed to match the observed  abundance of  QSO at z~6 remains highly uncertain (e.g. \cite{2014MNRAS.442.2036D}), it is difficult to determine  the required  RMS of the magnetic field.  However, it would be smaller than 0.1~nG and therefore is consistent with all the constraints.

The key assumption in our formalism is that magnetic flux conservation holds and so the magnetic field is frozen in with the matter. Phenomena like ambipolar diffusion and ohmic diffusion can void this assumption, though it can be shown that the magnetic braking timescales for sufficiently strong magnetic fields are much shorter than the ambipolar diffusion timescales for relatively low densities and thus studying magnetic braking under the assumption of flux-frozen magnetic fields is justified under the above mentioned conditions \citep{1979ApJ...230..204M}.

We studied the possible dynamical role a primordial magnetic field might play in removing angular momentum in models of super-massive black hole formation in the high-redshift Universe. Our analytic model provides quantitative results  which suggest that the presence of a primordial magnetic field of strength $B_0 \gtrsim 0.1$~nG could provide the necessary magnetic braking, in agreement with the time scale findings of \citet{1995Princeton}. This magnetic field strength is compatible with the existing upper bounds on the primordial magnetic field and is strong enough to have seeded the observed galactic magnetic fields.

It is of significant interest to firm up our estimate of the effectiveness of cosmological magnetic braking, by going beyond the simple, tractable, analytical model of high-redshift super-massive cosmological black hole formation on which it is based.

\section*{Acknowledgement} 
This work was partially supported by DOE grant DE-SC0019038.

\bibliographystyle{mn2e}
\bibliography{references}




\end{document}